\documentclass[extendedabs]{recpad2k}

\usepackage{booktabs}

\title{An audio-only method for advertisement detection in broadcast television content} 

\addauthor{Ant\'{o}nio Ramires}{antonio.ramires@inesctec.pt}{1}
\addauthor{Diogo Cocharro}{diogo.m.cocharro@inesctec.pt}{1}
\addauthor{Matthew E. P. Davies}{mdavies@inesctec.pt}{1}

 \addinstitution{
   INESC TEC \\
   Sound and Music Computing Group\\
   Rua Dr. Roberto Frias, s/n\\ 
   4200 - 465 Porto, Portugal 
  }

\runninghead{Student, Prof, Collaborator}{RECPAD Author Guidelines}

\begin{document}

\maketitle

\sloppy

\begin{abstract}
We address the task of advertisement detection in broadcast television content. While typically approached from a video-only or audio-visual perspective, we present an audio-only method. Our approach centres on the detection of short silences which exist at the boundaries between programming and advertising, as well as between the advertisements themselves. To identify advertising regions we first locate all points within the broadcast content with very low signal energy. Next, we use a multiple linear regression model to reject non-boundary silences based on features extracted from the local context immediately surrounding the silence. Finally, we determine the advertising regions based on the long-term grouping of detected boundary silences. When evaluated over a 26 hour annotated database covering national and commercial Portuguese television channels we obtain a Matthews correlation coefficient in excess of 0.87 and outperform a freely available audio-visual approach. 
\end{abstract}

\section{Introduction}
\label{sec:intro}

The classification of audiovisual content into categories and the identification of advertising has become increasingly important for end-users, broadcasters and entities that have contracted advertising space. This has special importance in the case of television content both for the need to archive content with the advertising removed, and in streaming contexts to allow for the region-specific substitution of advertising. 

Currently, the delimitation of advertising segments i.e., the identification of the beginning and end moments of a contiguous set of advertising content is typically performed by a human operator. As a result, the process is labour-intensive, expensive, and potentially error-prone \cite{conejero}. One means to improve the workflow of the human operator is to provide an automatic analysis of the broadcasting content which can classify the time-line into regions of advertising and regular programming. 

Existing algorithms for the detection of advertising fall into two main categories. Those which use explicit prior knowledge of a known set of advertisements and identify them using fingerprinting methods \cite{cardinal2010content}, and those which rely on heuristics as advertising indicators. For both types of approach, information can be leveraged from the video signal alone (logos, black frames, scene changes etc.), or in combination with the audio stream within audio-visual approaches \cite{covell2006}. 

In this work, our focus is on audio-only approach for television advertising detection which makes no use of video information, meta-data concerning the content type, or any prior knowledge of which advertisements can appear. To this end, we seek to discover if there is sufficient information in the audio signal alone to locate where advertisements occur. 
From this perspective, relevant acoustic cues include the presence of silence (often co-occurring with black frames at content boundaries), the presence of jingles, fast paced narration, background music, and identified repeated content -- which operates on the assumption that advertisements are repeated more frequently than regular programming. 

In our audio-based approach, we focus on a single acoustic property, that of silence -- which we assume to indicate very low signal energy rather than digital zeros in the audio bit-stream. We believe that silences have been under-used in the existing literature having been treated as just one feature among many which contribute towards the final decision. In our approach, we seek to maximize the information that can be obtained from detecting silences. Furthermore, we propose that by effective characterisation of different types of silences and the large scale grouping of an identified set of ``boundary silences'' we can obtain a very reliable descriptor of advertising boundaries in television.    

\section{Approach}
Our approach centres on the existence and detection of short pauses of silence (i.e., very low audio signal energy) in between separate pieces of content. We now provide an overview of each stage of the algorithm. Throughout, we assume the audio signal (a stereo signal sampled at 48kHz with 24-bit precision) has already been separated from the video content, and mixed down to mono. We notate the audio input as, $x$.   

\begin{figure}[ht]
  \centering
  \includegraphics[width=0.85\columnwidth]{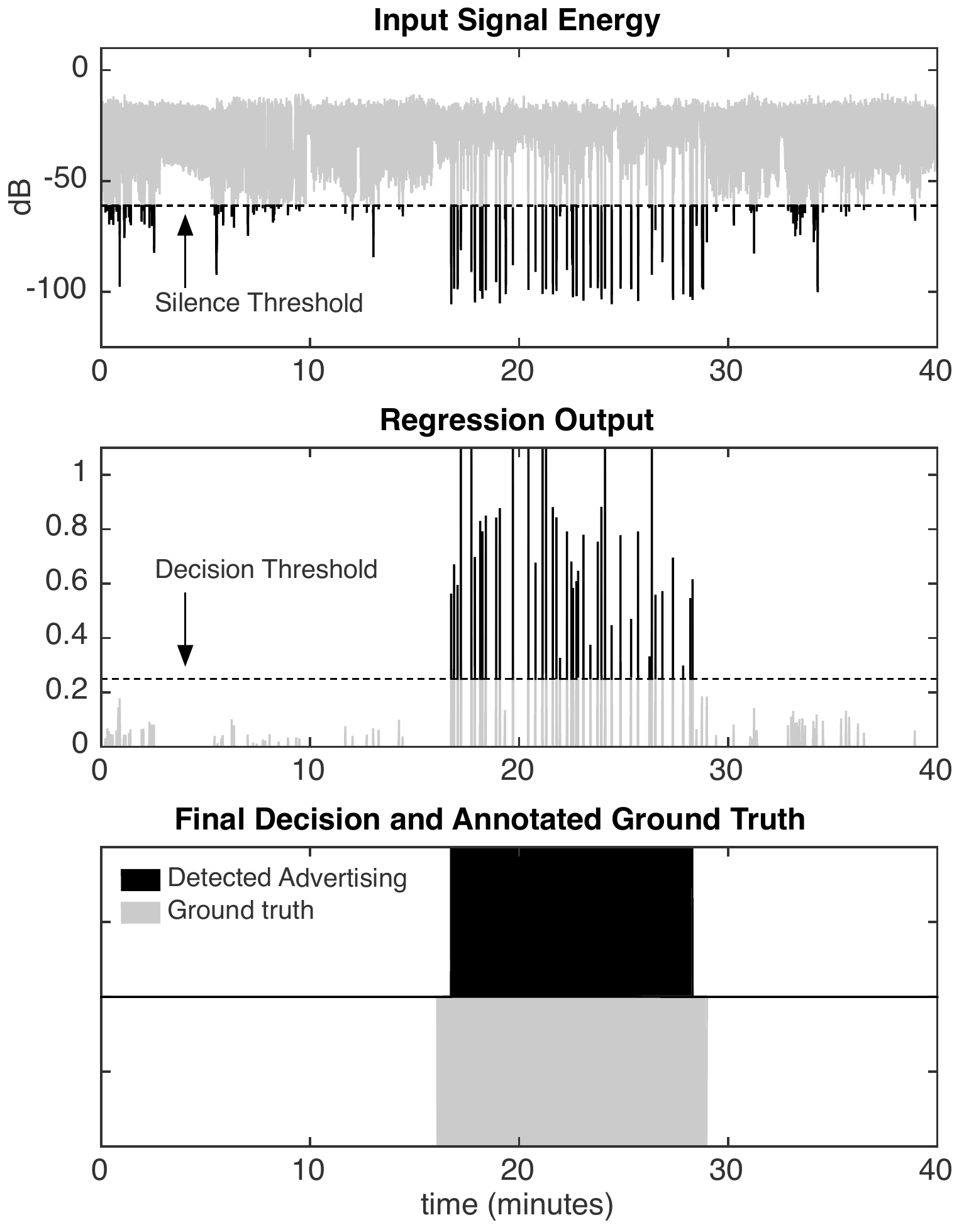}
  \caption{(top) Energy of input signal, with regions under the silence threshold shown in black. (middle) Output regression model on detected silences. Points above the decision threshold are shown in black. (bottom) The output classified as advertising and the corresponding ground truth.}
  \label{fig:1}
\end{figure}

To maintain parity with video frame rate of the television content (and allow easy integration with future video-based analysis) we partition $x$ into non-overlapping audio frames of 1920 samples (equivalent to 25 video frames per second). In each audio frame, $x_i$, we calculate the signal energy, 
$e_i = 20 \log_{10} \left( \sqrt{\textrm{mean}( {x_i}^2 )} \right).$
By taking the measurement in dB, we force all low energy parts of the signal to take large negative values. Next, to find all the low energy points in the input signal, we compare $e_i$ at each frame, $i$ to a silence threshold, $\eta$=-60\,dB, and retain those frames $i_s$ for which $e_i \le \eta$. An example is shown in the top plot of Fig. \ref{fig:1}. 

\begin{figure}[ht]
  \centering
  \includegraphics[width=0.85\columnwidth]{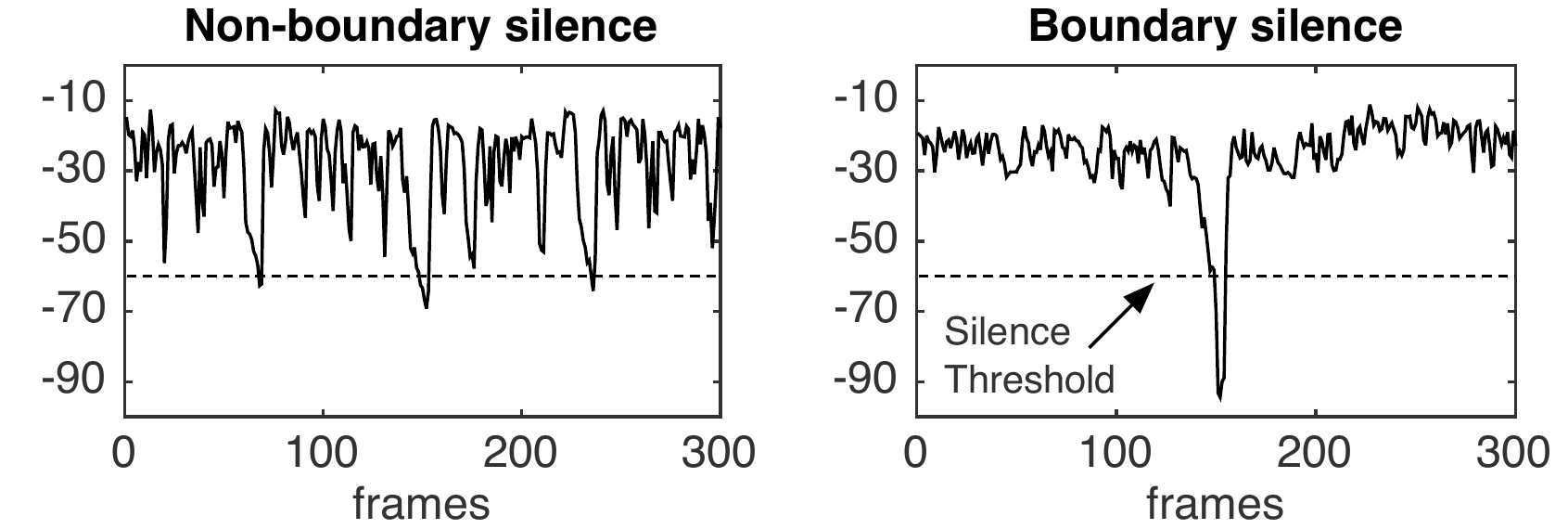}
  \caption{Examples of non-boundary silence (left) and boundary silence (right). The detected silence is at the mid-point of each plot.}
  \label{fig:2}
\end{figure}

Since short regions of silence can occur naturally within programming, e.g. as pauses between speech (either during narration or interviews with no background music or noise), we must filter out those silences which do not correspond to content boundaries. In our model we assume a boundary silence to be: short in duration, have a low minimum value, and be surrounded by regions of much higher energy. From a broadcast perspective we understand this is perceptually loud advertising content either side of a brief, imperceptible drop in energy, as shown in Fig. \ref{fig:2}.  

To distinguish between different types of silence we collect a small set of statistics: the max, mean, min, inter-quartile range, standard deviation, skewness, and kurtosis from a small temporal window of $\pm$6\,s ($\pm$ 150 audio frames) of the energy signal $e$ surrounding each detected silence $i_s$. We then perform a basic multiple linear regression on the extracted features where positive examples (i.e. annotated boundary silences) are labelled as 1, and non-boundary silences are labelled as 0. The output of the regression is shown in the middle plot of Fig. \ref{fig:1}. Here, all detected silences greater than the decision threshold, $\beta$=0.25, are retained and set to a value of 1, with all others discarded.

In the final stage of our algorithm, we pass a sliding rectangular window of 150\,s duration across the thresholded regression output. We determine regions of advertising as those which adhere to the following two conditions: i) there is more than one detected boundary silence within the long-term window (i.e. at least one starting and one ending silence); ii) the total duration of any period of advertising must be at least 60\,s. In this way isolated silences or those which are far from one another are excluded. The start of the detected advertising region is marked at the frame where the first detected boundary silence exits the long-term window. Likewise the end of the region occurs at the frame when the final boundary silence of any group exits the long-term window. An example of final output of the system is shown in the bottom plot of Fig. \ref{fig:1}.

\section{Results and Discussion}
We evaluate our algorithm over an annotated dataset we have compiled covering national (two instances of RTP 1 and one of RTP 2) and commercial channels (SIC and TVI) of Portuguese television. The dataset contains over 26 hours of content (segmented in 28 programmes), which has been annotated at two levels. First, to mark the high level boundaries between regular programming and advertising blocks, and second at a finer temporal level to marks the boundaries between all commercials. We use this second level for training the linear regression model. 

In order to measure the performance of silence-based method, we first count the number of true positives, $T_P$, true negatives, $T_N$, false positives, $F_P$, and false negatives, $F_N$, where a $T_P$ corresponds to a region which is both annotated and detected as advertising. 

As we can expect with broadcast television content, a far greater proportion of the content corresponds to scheduled programming rather than advertising (in our case, approximately 12\% is advertising). While many approaches in the literature report the F-measure as a performance indicator for advertising, this excludes any information about the number of $T_N$. To incorporate this information we instead report Matthews correlation coefficient, $M$, which is calculated as follows: 
\begin{equation}
M = \frac{T_P \times T_N - F_P \times F_N }{\sqrt{(T_P+F_P)(T_P+F_N)(T_N+F_P)(T_N+F_N)}}
\end{equation}
In addition to reporting the performance of our proposed approach, we also ran an open source audio-visual approach called ComSkip\footnote{\url{http://www.kaashoek.com/comskip/}, v. 0.82, accessed 06-15-2017.} under the default parameter settings. 
A comparison of performance between the two approaches is shown in Table \ref{tab:1}.
\begin{table}[t]
\centering 
\caption{Summary of dataset and comparison of algorithm performance.}
\label{tab:1}
\begin{tabular}{l c c c c }
\toprule 
Input                   & Total     & Advertising   &   ComSkip            & Proposed Alg.        \\
Channel                 & Duration  & Duration      &   Accuracy     & Accuracy  \\
    \midrule
$ \textrm{RTP 1}_a $    & 6h52m     & 0h23m        &   0.426       &    0.782\\
$ \textrm{RTP 1}_b $    & 1h10m     & 0h11m        &   0.007       &    0.863\\
RTP 2                   & 8h25m     & 0h24m        &  0.366        &    0.499\\
SIC                     & 8h37m     & 2h18m       &  0.648        &    0.931\\
TVI                     & 1h13m     & 0h9m         &   0.794       &    0.966\\
    \midrule
Overall                 & 26h17m    & 3h24m       &  0.610   &    \textbf{0.874}\\
    \bottomrule
\end{tabular}
\end{table}

As can be seen, our proposed approach outperforms ComSkip across all channels, with a correlation coefficient in excess of 0.87. Indeed, our approach performs especially well on the commercial channels (SIC and TVI), which contain large blocks of advertising content (running into several minutes at a time) with explicit use of silences between individual advertisements. 

The lowest performance was obtained on RTP 2. This channel contained a far lower proportion of commercial advertising, with the breaks between programming more frequently containing trailers for upcoming in-channel content (and without such prominent silence boundaries). Since this content falls between the main programming, it can be understood as advertising, and thus something which our current approach cannot readily detect. However, given the critical requirement in advertising removal applications not to misclassify programming as advertising, our proposed approach has explicitly been parameterised to minimise false positives. To this end, it provides ``conservative'' estimates of advertising boundaries. Indeed, over the 26 hours, our approach has just 6 false positive frames, with ComSkip having only 761 false positive frames ($\sim$ 30\,s).    

A potential criticism of the comparative results is that they may be somewhat optimistic since our approach has partial access to the dataset for training, where as ComSkip does not. However, our multiple linear regression model was trained using leave one out cross fold validation at the programme level, and therefore we maintain some separation between training and testing material. Informal tests on currently un-annotated validation data also indicates highly promising performance and larger-scale evaluation will be among the main areas of future work.     

\section{Conclusions}
We have a presented a new audio-only approach for the detection of advertising in television broadcast content. Our approaches relies on the short, medium, and long-term modelling of silences within the audio stream as a means for distinguishing regular programming from advertising. A novel feature of our approach is the ability to reject silences (e.g. pauses in speech) which do not exhibit the statistical properties of content boundaries. Currently our approach has been optimised for Portuguese television content, therefore main focus of our future work will be to investigate the accuracy of our approach on international television content. Furthermore, we intend to enhance our audio-only model via the inclusion of other important cues includes in jingle detection, music/speech separation and audio production effects related to bandwidth and dynamic range. 

\section{Acknowledgements}
This article is a result of the project MOG CLOUD SETUP - N\textsuperscript{o}17561, supported by Norte Portugal Regional Operational Programme (NORTE 2020), under the PORTUGAL 2020 Partnership Agreement, through the European Regional Development Fund (ERDF).

Diogo Cocharro was supported by Project TEC4Growth-Pervasive Intelligence, Enhancers and Proofs of Concept with Industrial Impact/NORTE-01-0145-FEDER-000020 is financed by the North Portugal Regional Operational Programme (NORTE 2020), under the PORTUGAL 2020 Partnership Agreement, and through the European Regional Development Fund (ERDF). 



\bibliography{eg}

\begin{thebibliography}{3}
\providecommand{\natexlab}[1]{#1}
\providecommand{\url}[1]{\texttt{#1}}
\expandafter\ifx\csname urlstyle\endcsname\relax
  \providecommand{\doi}[1]{doi: #1}\else
  \providecommand{\doi}{doi: \begingroup \urlstyle{rm}\Url}\fi

\bibitem[Cardinal et~al.(2010)Cardinal, Gupta, and
  Boulianne]{cardinal2010content}
P.~Cardinal, V.~Gupta, and G.~Boulianne.
\newblock Content-based advertisement detection.
\newblock In \emph{{INTERSPEECH}}, pages 2214--2217, 2010.

\bibitem[Conejero and Anguera(2008)]{conejero}
D.~Conejero and X.~Anguera.
\newblock {TV} advertisements detection and clustering based on acoustic
  information.
\newblock In \emph{Intl. Conf. on Computational Intelligence for Modelling
  Control Automation}, pages 452--457, 2008.

\bibitem[Covell et~al.(2006)Covell, Baluja, and Fink]{covell2006}
M.~Covell, S.~Baluja, and M.~Fink.
\newblock Advertisement detection and replacement using acoustic and visual
  repetition.
\newblock In \emph{IEEE Workshop on Multimedia Signal Processing}, pages
  461--466, 2006.

\end{thebibliography}

\end{document}